\documentclass[pre,twocolumn,amsmath,amssymb,floatfix]{revtex4-1}
\usepackage{graphicx}% Include figure files
\usepackage{epsfig}% Include figure files
\usepackage{dcolumn}% Align table columns on decimal point
\usepackage{bm}% bold math
\usepackage{color}
\usepackage{titlesec}

\def\(({\left(}
\def\)){\right)}
\def\[[{\left[}
\def\]]{\right]}

\newcommand{\beq}{\begin{equation}}
\newcommand{\eeq}{\end{equation}}
\newcommand{\barr}{\begin{eqnarray}}
\newcommand{\earr}{\end{eqnarray}}
\newcommand{\bei}{\begin{itemize}}
\newcommand{\eei}{\end{itemize}}

\begin{document}

\title{Finite temperature crossovers in periodic disordered systems}

\author{L.~Foini}
\affiliation{Department of Quantum Matter Physics, University of Geneva, 24 Quai Ernest-Ansermet, CH-1211 Geneva, Switzerland}
\author{T.~Giamarchi}
\affiliation{Department of Quantum Matter Physics, University of Geneva, 24 Quai Ernest-Ansermet, CH-1211 Geneva, Switzerland}

\date{\today}

\begin{abstract}
We consider the static properties of periodic structures in weak random disorder.
We apply a functional renormalization group approach (FRG) and a Gaussian variational
method (GVM) to study their displacement correlations.  We focus in particular on the effects of temperature and we compute explicitly the crossover length scales separating
different regimes in the displacement correlation function. To do so using the FRG we introduce a functional form that approximate very accurately the flow of the disorder correlator at all scales. We compare the FRG and GVM results and find excellent agreement. We show that the FRG predicts in addition the existence of a third length scale associated with the screening of the disorder by thermal fluctuations and discuss a protocol to observe it. 
\end{abstract}

\maketitle

\section{Introduction}

Understanding the properties of elastic systems in disordered environment
represents a key problem in physics because of their relevance in a number of experimental
situations and their own theoretical interest.
Despite very different microscopic mechanisms, a large variety of systems can be described as elastic manifold embedded in
random media \cite{G09}.
Typically these are divided into two categories.
One encompasses interfaces in magnetic~\cite{LFCMGLD98,GBFJKG14}, ferroelectric~\cite{TPGT02,PGT05} materials or spintronic systems \cite{YIMBMO07}, fluid invasion
in porous media~\cite{WW83} and fractures~\cite{BBFRR02,BSP08}. The second concerns random periodic systems such as
charge density waves~\cite{BN04}, vortex lattices in type II superconductors~\cite{BFGLV94} and Wigner crystals~\cite{CGNS05l}.
All these systems are characterized by the competition between an elastic energy that
wants the manifold flat or the periodic system ordered and the impurities -- that are inevitably present in any real system -- that tend to distort it in
order to accommodate it in the optimal positions.
This competition results in a number of interesting physical features
ranging from self-similarity in their static correlation functions to
a very rich (and glassy) dynamical behavior~\cite{ALG12}.

While the static asymptotic properties of both interfaces and periodic systems
are well understood
at present at very low temperature, the effects of temperature are still unclear in most systems. In particular for
interfaces this questions has recently been the focus of several studies (see e.g. \cite{DGGB10,ALG10,ALG13a,BLDR10} and refs therein).

The corresponding question in the second class of systems, i.e.
in periodic structures, is still largely not explored.
Despite the similarities in the theoretical modeling,
periodic systems show some important differences compared to interfaces,
in particular for weak disorder quasi-long range positional order exists~\cite{GLD94,GLD95,GKR06}, at variance with the power-law roughening of interfaces.
In most of the analyses on such systems the effect of temperature has been mostly disregarded since they are controlled by a zero temperature fixed point and thus
low temperatures are not essentially affecting the asymptotic behavior
of the correlation functions with distance.
However, as is the case for interfaces, temperature can affect both the amplitude of asymptotic regimes and generate crossover scales and intermediate distance regimes.
It is thus interesting, especially in view of contact with experimental systems to have a better understanding of such effects.

Two methods that have been employed with great success for the study of periodic systems are the functional renormalisation group
(FRG) method and a Gaussian variational method (GVM). Initially introduced for the interfaces~\cite{MP91,F86}, they have been extended
to deal with periodic systems as well~\cite{GLD94,GLD95} and shown to give consistent results to each other.
Static correlation functions have been computed using these methods~\cite{K93,GLD94,BEN01}.
However for the FRG the zero temperature fixed point was assumed from the start, so the relevant length scales created by the finite temperature were not investigated.

In this paper we fully incorporate the temperature effects in the FRG and use this technique to investigate the various scales
that are created by the finite temperature in the displacement correlation functions. We compare these results with the ones obtained by the GVM and we show
that they give consistent results, concerning the different length scales characterizing the relative displacement correlation functions of the system.
On a more technical level since incorporating the finite temperature effects in the FRG leads to quite complicated equations
we also show that there is an efficient fitting scheme that approximate very well the disorder correlator all along the FRG
flow and use it for the present problem. Such fitting form is also particularly useful in the more complicated case of the dynamics
of such disordered systems at finite temperature. Such analysis will  be presented elsewhere~\cite{FGpreparation}.

The paper is organized as follows: in section~\ref{Sec_model} we introduce the model for periodic disordered elastic systems;
in section~\ref{Sec_FRG}  we outline the strategy to study the problem
by the functional renormalization group technique and  in section \ref{Sec_FRG_fitting} we introduce a fitting form
to the correlator entering in the FRG equations and use it to solve the equations.
In section~\ref{Sec_FRG_Sq} we explain the way the Fourier transform of the displacement correlation function (FTD) is computed by FRG and
in section~\ref{Sec_FRG_lengths} we discuss the length scales that characterize the
displacement correlation function obtained by FRG.
In section \ref{Sec_Variational} we use a variational approach to obtain the crossover lengths.

%outline the variational approach, in section \ref{GVM_RSB} we detail the solution obtained
%with a replica symmetry breaking ansatz and
%in section \ref{Sec_GVM_Sq} we derive the displacement correlations by GVM.
%In section~\ref{Sec_GVM_lengths} we highlight the length scales that appear from the structure factor obtained by GVM.

In section \ref{Sec_Sq_Comparison} we compare the results for the exponents governing the FTD
and the different length scales obtained within the two approaches and discuss their salient features. In section \ref{Sec_concl} we conclude.

\section{Model}\label{Sec_model}

\begin{figure}
  \begin{center}
    \includegraphics[height=6cm]{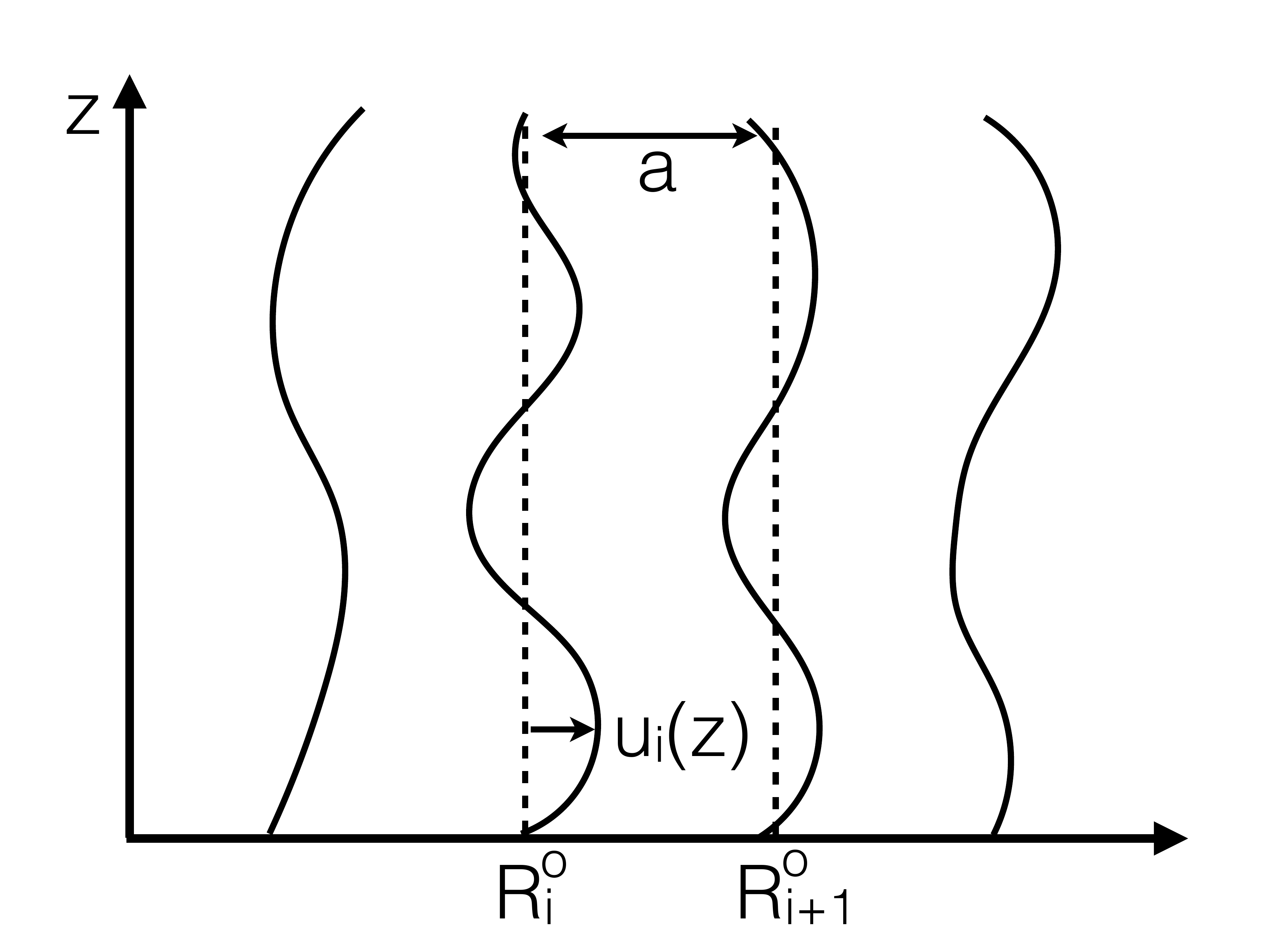}
        \caption{A set of lines forming a periodic lattice in a plane, put in a random environment can be described
        in the elastic limit by the Hamiltonian (\ref{H}).}
    \label{Fig:lines}
  \end{center}
\end{figure}

We consider a periodic elastic system
where the position of each particle is characterized by a coordinate
$R_i = R_i^o + u_i$ and $R_i^o$ forms a perfect lattice.
$u_i$ is the displacement field that we consider in the elastic limit $u_{i+1}-u_i \ll a$
where $a$ is the typical lattice spacing. In this case $u_i$ can be replaced by
a continuous field and the energy of the system in a disordered environment can be approximated
by the Hamiltonian:
\beq\label{H}
H = \frac{c}{2} \int {\rm d}^d r  \ | \nabla u|^2 +  \int {\rm d}^d r \ W(r) \rho(r,u)
\eeq
where $c$ is the elastic constant. For simplicity we have taken the displacements $u_i$ as scalars and thus considered here
only a single elastic constant. The extension to more complex elastic forms is straightforward (see e.g.~\cite{GLD95}).
The Hamiltonian~(\ref{H}) would for example describe a set of lines constrained to move in planes in a two-dimensional or three-dimensional lattice (see Fig.~\ref{Fig:lines} for the two dimensional version).
The density $\rho$ can be expressed~\cite{GLD95} using the vectors of the reciprocal lattice $K$, $\rho = \rho_0 \sum_{K\neq 0} e^{i K (x - u(r))}$
where $\rho_0$ is the average density and $x$ is the coordinate along the direction of $u$.
If we take the random potential $W(r)$ to be Gaussian and with correlations $\overline{W(r)W(r')} = \delta(r-r') D_0$
the effective potential in the Hamiltonian (\ref{H}) reads $V(r,u) =  W(r) \rho(r,u)$
with correlations $\overline{V(r,u)V(r',u')} = R(u-u')\delta(r-r')$ and $R(u-u') = D_0 \rho_0^2 \sum_{K\neq 0} e^{i K (u(r)-u'(r))}$
where we disregarded rapidly oscillating terms.
In the following for simplicity we will assume $R(u)$ as made of a single harmonic, namely:
\beq\label{Ru}
R(u) = D \cos(K u)
\eeq
and we will work with the correlator of the random force $F_{dis}$ such that
$\overline{F_{dis}(r,u)F_{dis}(r',u')} = \delta(r-r') \Delta(u-u')$ with $\Delta(u-u')=-R''(u-u')$. Such a situation is for example pertinent for charge density waves~\cite{BN04}.

We identify the roughness exponent $\zeta$ as the one entering in
the correlation of the displacement field $\langle (u(r)-u(0))^2 \rangle = |r|^{2 \zeta}$.
Various regimes can be identified in the relative displacement correlation function.
In particular at zero temperature systems with a single harmonic exhibit two
regimes. depending on the length scale: a Larkin regime where $\zeta=(4-d)/2$ at
the smallest scales~\cite{L70} crossing over to the random periodic phase asymptotically with $\zeta=0$ and logarithmic grow of the displacements $\zeta=0$. In presence of several harmonic a third regime (random manifold) would exist of the displacements~\cite{GLD95}.
We will concentrate here on the simple case of the single harmonic to focus on the additional length scales appearing with the temperature.
We also consider that the elastic model is valid at all temperatures, i.e. that no topological defects will appear in the system. Such an assumption is exact for the above system of lines.

\section{Functional RG approach}\label{Sec_FRG}

We define $\tilde{\Delta}(u)= \frac{S_d \Lambda^d}{(c \Lambda^2)^2} \Delta(u)$
and $\tilde{T} = \frac{S_d \Lambda^d}{c \Lambda^2} T$ where $S_d$ is the surface
of the hypersphere in $d$ dimension divided by $(2 \pi)^d$ and $\Lambda$ is
an ultraviolet cutoff.

Upon variation of the cutoff the correlator of the disorder and the other physical quantities are renormalized. The main difficulty of such disorder systems is that the
whole function should be kept, leading to a functional renormalization.
The FRG equation governing the statics of the correlator of the  force specialized to random periodic (RP) systems
having a roughness exponent $\zeta=0$ reads~\cite{F86,GLD95,CGLD00}:
\beq\label{Flow_RP_v0}
\begin{array}{ll}
\displaystyle \partial_l \tilde{\Delta}(u) & =
\displaystyle\epsilon \tilde{\Delta}(u)
+ \tilde{T} \tilde{\Delta}''(u)
\\ \vspace{-0.2cm} \\
&\displaystyle \qquad\qquad+ \tilde{\Delta}''(u) [\tilde{\Delta}(0) - \tilde{\Delta}(u) ]
- (\tilde{\Delta}')^2
\\ \vspace{-0.2cm} \\
& \displaystyle \partial_l \tilde{T} \displaystyle = (\epsilon-2)  \tilde{T} \ .
   \end{array}
\eeq
At zero temperature this equation is known to lead to a singularity around the origin $\tilde{\Delta}(0)$
after a finite length scale $l_c$ in the flow.
In particular the static length scale at which the curvature of the correlator $\Delta''(0)$
blows up for $T=0$ is defined as~\cite{F86,CGLD00}: 
\beq\label{Eq_length_lc}
l_c = \frac{1}{\epsilon} \log \Big( 1 + \frac{\epsilon}{3 \tilde\Delta_0''(0)} \Big)  \ .
\eeq
the so called Larkin length.
The presence of temperature in the flow (\ref{Flow_RP_v0}) cures this non analyticity  rounding the singularity (cusp in the function $\Delta$).
For $d>2$ the cusp around the origin in presence of temperature appears asymptotically at large scales for which the temperature renormalizes to zero according to (\ref{Flow_RP_v0}).
The fixed point solution for RP systems for $u \in [0,1]$ is known exactly and reads~\cite{GLD95,CGLD00}:
\beq\label{Eq_fixed_point}
\tilde\Delta^{\ast}(a u) = \frac{a^2 \epsilon}{6} \Big( \frac{1}{6} - u (1-u) \Big) \ .
\eeq
The function is continued periodically for $u \neq [0,1]$
and the non-analyticity around $u=0$ is evident.

In order to study the effects of finite temperature we need not only the fixed point but the full flow. To study numerically the flow we start the procedure with a correlator
of the form $\tilde{\Delta}(u)=\tilde{\Delta}_0 \cos(2\pi u)$ with
$\tilde{\Delta}_0=0.005$ and we focus on the flow in the interval $u\in [-0.5,0.5]$.
We discretize this domain in $2N+1$ intervals with $N=1000$. The discretization
in the running length of the flow is set to $\delta l = 10^{-6}$.
The flow is then obtained by solving the differential equations
using a finite difference method.
We used forward first derivative for $u<0$ and backward first derivative
for $u>0$. The point $u=0$ was treated with a central first derivative until
its second derivative reaches a threshold value beyond which it was
taken a forward derivative. Second order derivatives were considered all central.

\begin{figure}
  \begin{center}
    \includegraphics[height=6cm]{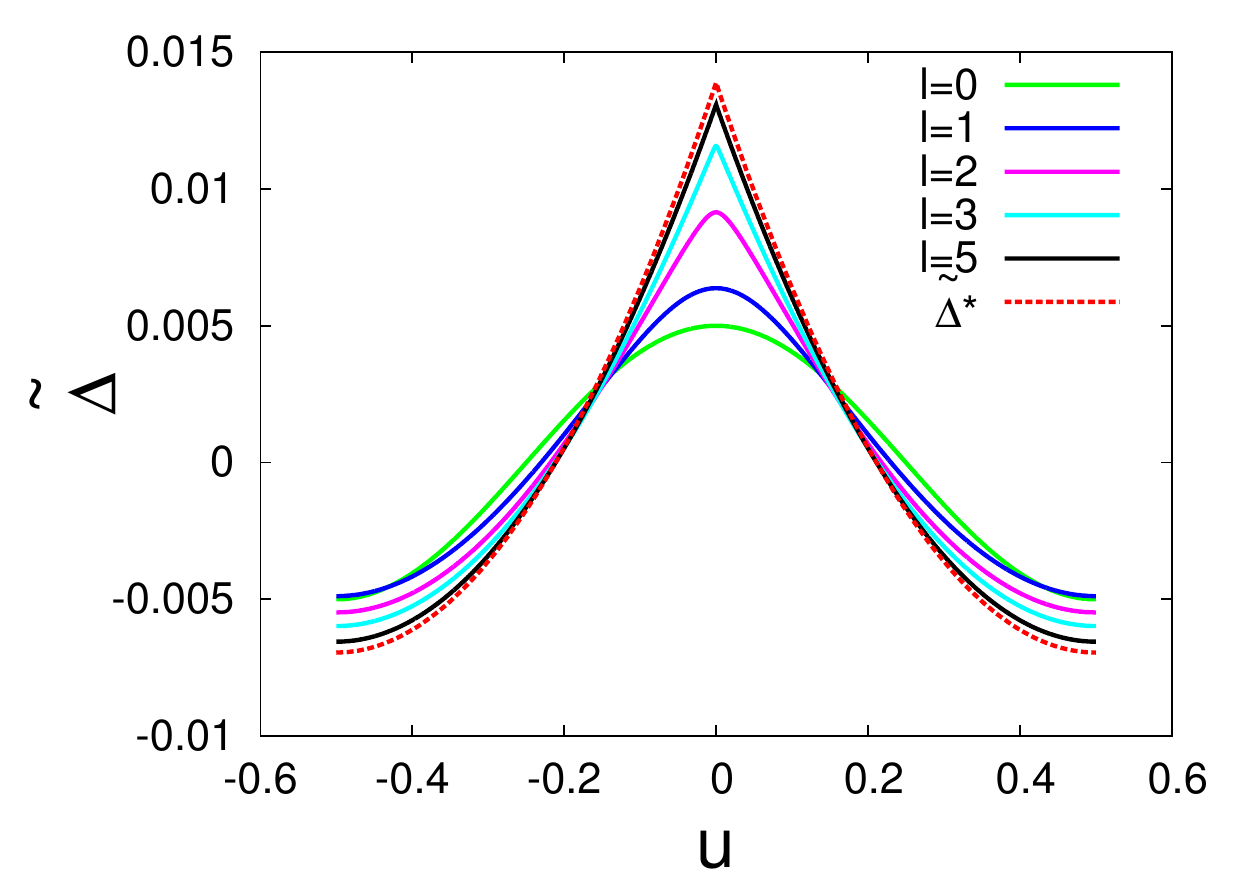}
        \includegraphics[height=6cm]{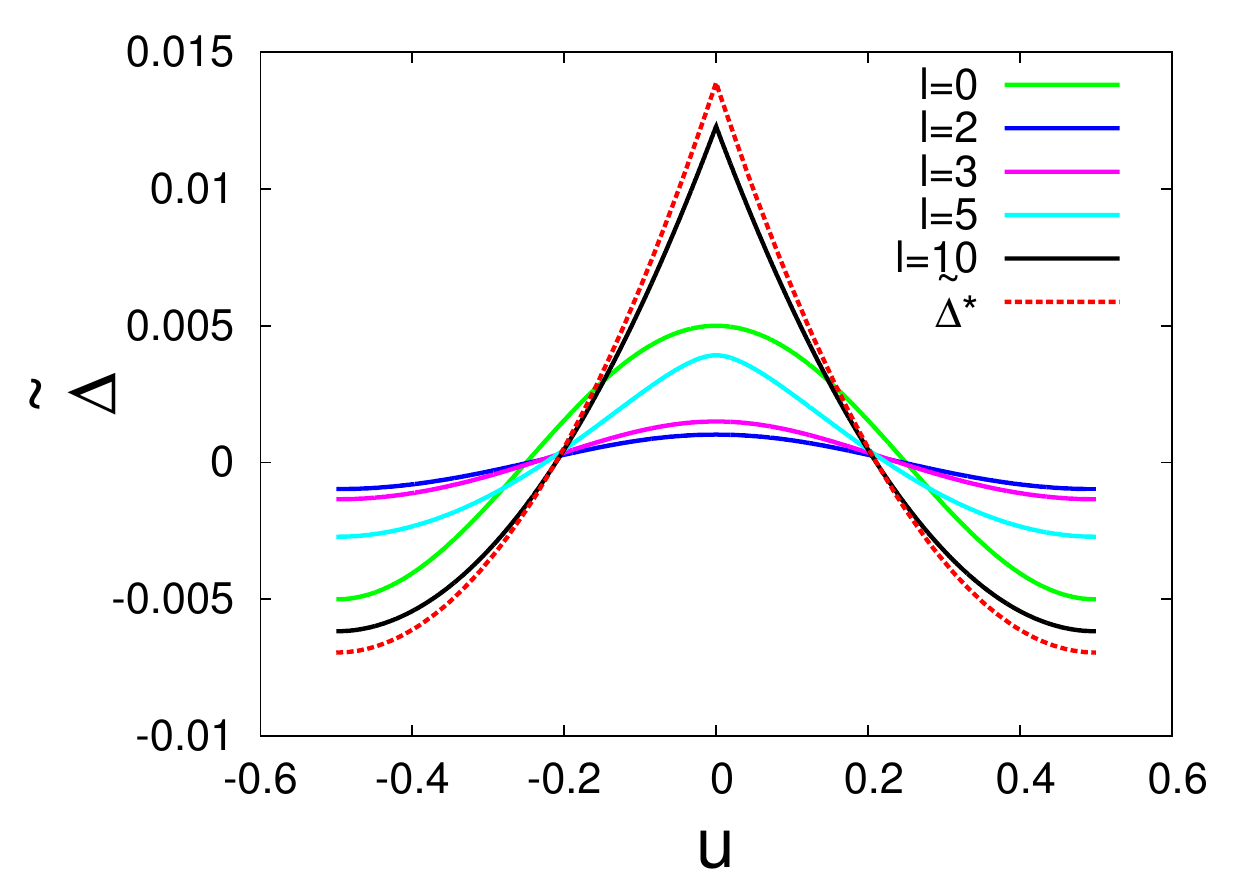}
        \caption{The disorder correlator at different scales with two different initial temperatures:
       $\tilde{T}=0.01$  in the upper panel and $\tilde{T}=0.1$ in the lower panel. The flow in the two cases is different.
       In the upper panel the flow is ``monotonically" towards the asymptotic value $\tilde{\Delta}^{\ast}$.
       In the lower panel the flow is first towards
       a vanishing amplitude function and then, only after a finite length scale, the flow goes back to the asymptotic
       zero temperature function $\tilde{\Delta}^{\ast}$. }
    \label{Fig:FRG_T_v0}
  \end{center}
\end{figure}

In Fig.~\ref{Fig:FRG_T_v0} we show the behavior of the correlator $\tilde{\Delta}_l(u)$ under FRG with
the temperature initialized to the value $\tilde{T}=0.01$ (upper panel)
and $\tilde{T}=0.1$ (lower panel).
In the upper panel of Fig.~\ref{Fig:FRG_T_v0} we see that the flow tends ``monotonically" towards the fixed point.
In the lower panel of Fig.~\ref{Fig:FRG_T_v0} instead,
the correlator first flows towards a vanishing amplitude function and after a certain length scale
flows towards the fixed point $\tilde{\Delta}^{\ast}$, shown with a red dashed line.

\subsection{Fitting form}\label{Sec_FRG_fitting}

Solving the full equations although potentially feasible is quite taxing. We show in this section that an excellent approximation of the solution can be obtained by a fitting function depending on only few parameters.
\begin{figure}
  \begin{center}
          \includegraphics[height=6cm]{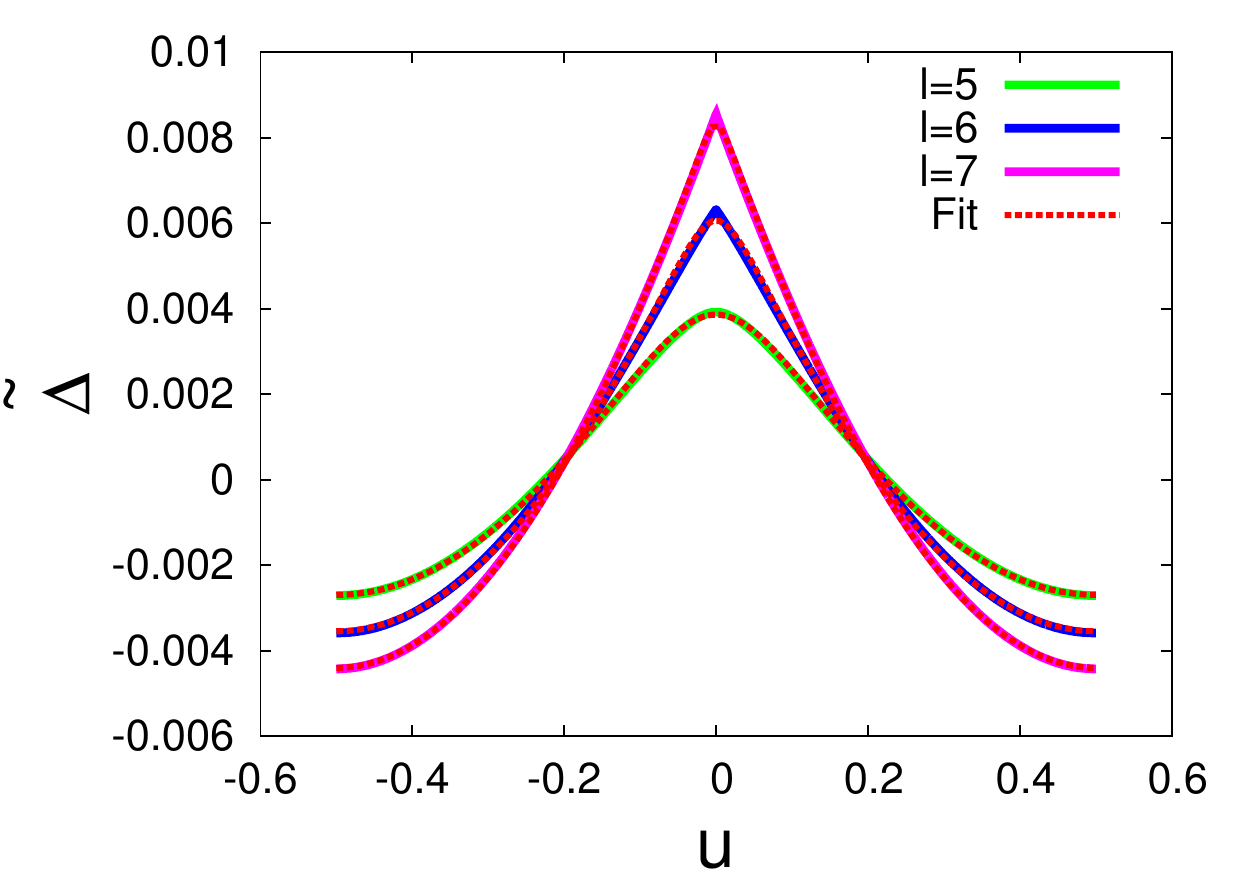}
        \caption{With solid lines we show  the disorder correlator obtained by numerical integration of Eq.~(\ref{Flow_RP_v0})
        with $\tilde{T}=0.1$, $\epsilon=0.5$ and $\tilde{\Delta}_0=0.005$,
        at three different lengths, $l=5$ (green line), $l=6$ (blue line). and $l=7$ (pink line). With dashed red lines
        we show the fit to such curves obtained using the function~(\ref{Trial_function2}).
      In general the fit is always remarkably good, apart from a narrow interval of lengths where it is not accurate around the origin $u=0$. For the
        values  of $T$ and $\epsilon$ here shown the disagreement is around $l\simeq6$.}\label{Figs_Plot_trial_function}
  \end{center}
\end{figure}

We note that the following function:
 \beq\label{Trial_function}
\displaystyle \hat{\Delta}_l(u) = \phi_l - \sqrt{\alpha_l + \beta_l u^2} + \gamma_l u^2 \ ,
\eeq
always provides a very accurate fit of the function $\tilde{\Delta}$
at any scale (see Fig.~\ref{Figs_Plot_trial_function}). Typically
there is only a tiny interval in $l$ where one can see discrepancy
around the origin $u=0$ between the function $\tilde{\Delta}_l(u)$
and the fitted function~(\ref{Trial_function}) (see Fig.~\ref{Figs_Plot_trial_function}).
One can keep 4 free parameters in Eq.~(\ref{Trial_function}), but it might
be convenient (basically without changing the accuracy of the fit) to fix
the coefficient $\gamma_l $ (or equivalently $\alpha_l$ or $\beta_l$) in order to
ensure $\hat{\Delta}'(\pm \frac12)=0$ (and thus periodicity).
This choice leads to the following  constraint:
\beq
\displaystyle\gamma_l = \frac{\beta_l}{\sqrt{4 \alpha_l + \beta_l}} \ .
\eeq
In this case one is therefore interested in the flow of the function:
\beq\label{Trial_function2}
\displaystyle \hat{\Delta}_l(u) = \phi_l - \sqrt{\alpha_l + \beta_l u^2} + \frac{\beta_l}{\sqrt{4 \alpha_l + \beta_l}} u^2
\eeq
and in particular in the flow of the parameters $\alpha_l$, $\beta_l$ and $\phi_l$.
Having a good fitting function is particularly helpful for studying the dynamics in
presence of a non-zero velocity~\cite{FGpreparation}. In this case in fact computing correlation functions requires the knowledge of the whole correlator (and not just its value in $u=0$ as for the statics) and being able of representing the entire function only through three parameters greatly simplifies the analysis.

\subsection{Displacement correlation function within the FRG}\label{Sec_FRG_Sq}

We can now compute the Fourier transform of the displacement correlation function (FTD):
\beq
\Gamma(q) = T \tilde{G}(q) = \langle u(q) u(-q) \rangle
\eeq
which satisfies the RG flow equation (specialized to the case $\zeta=0$):
\beq\label{Eq_RG_Gamma}
\Gamma(q,T,\Delta) = e^{d l} \Gamma(q e^l,T e^{(2-d)l},\Delta(l)) \ .
\eeq
One can set $e^{l^{\ast}} q = 1/a$ in (\ref{Eq_RG_Gamma}) and obtain:
\beq\label{Eq_Gamma}
\Gamma(q,T,\Delta) = \Big(\frac{1}{q a}\Big)^d \Big[\frac{T_{l^{\ast}}}{c k^2} +  \frac{\Delta_{l^{\ast}}(0)}{c^2 k^4}\Big]_{k=1/a} \sim q^{-\nu}\ .
\eeq
The expression between squared brackets is the perturbative result
at finite temperature, as obtained within the Larkin model~\cite{L70}, which is valid
at the length scale $k^{-1}=a$.
The formula defines the exponent $\nu$ which is directly related to the
roughness exponent $\zeta$ via $\zeta=(\nu-d)/2$.

\begin{figure}
  \begin{center}
        \includegraphics[height=6cm]{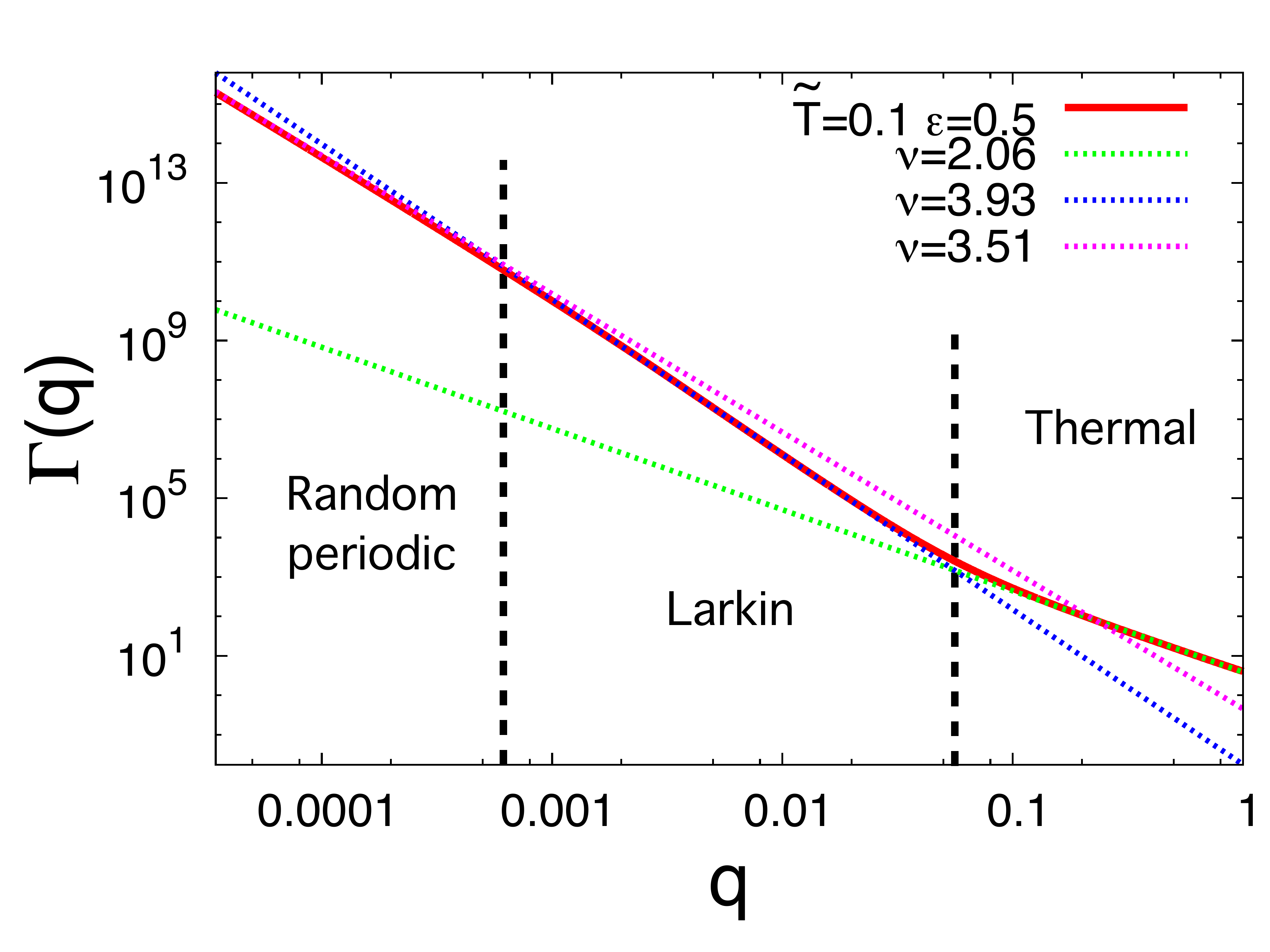}
         \caption{With a solid red line we show the displacement correlation function, as obtained from Eq.~(\ref{Eq_Gamma})
             with $\tilde{T}=0.1$ and $\epsilon=0.5$. Dashed lines highlights the
             different regimes characterized by different exponents. In particular, at large $q$ one finds
             the thermal regime with $\nu\simeq 2$ (green dashed line), at intermediate
             $q$ the Larkin regime (blue dashed line) with $\nu\simeq 4$ and
             at small $q$ the asymptotic exponent typical of random periodic system, i.e. $\nu=d=3.5$  (pink dashed line).
              }  \label{Fig_Structure_factor}
  \end{center}
\end{figure}

In Fig.~\ref{Fig_Structure_factor} we show the results obtained for the FTD
at $\tilde{T}=0.1$ and $\epsilon=0.5$. The logarithmic plot with dashed tangents shows the three
regimes encountered as a function of $q$. At very short length scales temperature fluctuations dominate the correlation and disorder is unimportant. One has a thermal regime with an exponent $\nu=2$. One then crosses over to the Larkin regime with the exponent
$\zeta=(4-d)/2$, i.e. $\nu=4$,
(which from our plot gives an exponent $\nu\simeq3.9$).
Finally
at large scales one recovers the exponent $\nu=d=3.5$ which characterizes the random periodic systems whose asymptotic behavior is given by $\zeta = 0$ and logarithmic correlation functions.

These three regimes define two crossover scales that we compute in the next section. In addition to these length scales that could be expected on a physical basis, we will show that the FRG gives evidence of a third length scale.

\subsection{Crossover length scales within the FRG}\label{Sec_FRG_lengths}

We can identify the crossover length $l_{th} = q_{th}^{-1}$ that separates the thermal from the Larkin regime by considering only the linear flow of $\tilde{\Delta}$. Indeed in these two regimes the effective disorder remains small.

Let us us solve the linearized flow.
\beq\label{Eq_lin}
\displaystyle \partial_l \tilde{\Delta}(u)=
\epsilon \,\tilde{\Delta}(u)
+ \tilde{T}e^{- l (d- 2)} \tilde{\Delta}''(u) \ .
\eeq
Defining $D_l = \tilde{T} \int_0^l {\rm d} l' e^{-(d-2) l'} = \frac{\tilde{T}}{d-2}( 1 - e^{-(d-2) l})$ the solution is given by~\cite{BLDR10}:
\beq
 \tilde{\Delta}(u)= \frac{e^{\epsilon l}}{\sqrt{4 \pi D_l}} \int {\rm d}u' \, e^{- \frac{(u-u')^2}{4 D_l}} \tilde{\Delta}_{l=0}(u')
\eeq
and if we choose $\tilde{\Delta}_{l=0}(u) = \tilde{\Delta}_0\cos(K u)$ it gives:
\beq\label{Sol_lin}
\begin{array}{ll}
 \displaystyle \tilde{\Delta}(u) & \displaystyle=  \tilde{\Delta}_0 e^{\epsilon l}e^{-K^2 D_l}  \cos\left( K u \right)
\\ \vspace{-0.2cm} \\
 & \displaystyle\simeq \tilde{\Delta}_0 e^{\epsilon l}e^{-K^2 \frac{\tilde{T}}{d-2}}
  \cos\left( K u \right) \ .
   \end{array}
\eeq

Using the solution~(\ref{Sol_lin}), we can extract $q_{th}$
from Eq.~(\ref{Eq_Gamma}),
as the point where the thermal part of $\Gamma(q)$ equals the disordered term.
This gives:
\beq\label{qth_FRG}
 q_{th} =\sqrt{\frac{\Delta_0}{c T}} e^{-K^2 \frac{T S_d \Lambda^{d-2}}{2 c (d-2)}} = K \sqrt{\frac{\tilde{D} }{\tilde{c} T}} e^{-K^2 \frac{L_{l}^2}{4}}  \ ,
\eeq
where we introduced the Lindemann length
which measures the strength of thermal fluctuations:
\beq\label{lT}
 L_{l}^2 = \langle u^2 \rangle_T = 2 T \int \ \frac{{\rm d}^d q}{(2\pi)^d} \frac{1}{c q^2} \simeq \frac{2 T S_d}{c (d-2)} \Lambda^{d-2} \ .
\eeq
Here and in the following we defined $c=\tilde{c}/a^d$ and $D= \tilde{D}/a^d$ where $\tilde{c}$ and
$\tilde{D}$ have the dimension of an energy and the squared of an energy.

The Larkin length $l_L=q_L^{-1}$, which marks the end of the Larkin regime and the
passage towards the asymptotic random periodic regime, can be defined as the point where the non linear terms
in the flow of $\tilde{\Delta}$ become
important with respect to the linear terms. From this criterium one gets:
\beq\label{qL_FRG}
\begin{array}{ll}
\displaystyle q_L & \displaystyle \simeq \Big( \frac{\Delta_0 S_d K^2}{c^2 \epsilon} \Big)^{1 / \epsilon}  e^{-K^2 \frac{T S_d \Lambda^{d-2}}{\epsilon c (d-2)}}
\\ \vspace{-0.2cm} \\
& \displaystyle = \frac1a \Big( \frac{ \tilde{D} S_d (K a)^4}{ \tilde{c}^2 \epsilon} \Big)^{1 / \epsilon} e^{-K^2 \frac{L_{l}^2}{2\epsilon}} \  .
   \end{array}
\eeq

The criterium to have a Larkin regime becomes $q_{th} > q_{L}$. This criterium is always satisfied at low and high temperatures. However there might be some intermediate range of temperature where the criterium is not satisfied and the Larkin regime disappears.
We will come back on that point in  Section~\ref{Sec_Sq_Comparison}

Quite interestingly, in addition to the above length scales $q_{th}$ and $q_{L}$
an additional crossover length scale can be identified from the FRG.

Indeed in the flow an additional length scale can be defined by the scale at which
in the flow of $\tilde{\Delta}$ the linear term in the flow (\ref{Flow_RP_v0})
proportional to the temperature
dominates with respect to the one multiplying $\epsilon$. This defines:
\beq\label{qT_FRG}
 q_{T} \simeq  \frac1a \Big( \frac{\tilde{c} \ \epsilon }{T S_d  (K a)^2 }  \Big)^{\frac{1}{2 - \epsilon }} \simeq \frac{1}{K a^2} \sqrt{ \frac{\tilde{c} \ \epsilon }{T S_d }} \ .
\eeq

At high enough temperatures, the inverse length scale $q>q_T$ are associated with the flow
of the correlator that tends towards a vanishing amplitude function, as shown in Fig.~\ref{Fig:FRG_T_v0}
for $\tilde{T}=0.1$. This behavior, understood as an effective reduction of the influence of the disorder due
to thermal fluctuations upon increasing the length scale, is manifested also in the (disordered part of)
FTD, as we discuss in Section~\ref{Sec_Sq_Comparison} and we show in Fig.~\ref{Fig:Correction_Sq}.
The quantity~(\ref{qT_FRG}) can be made small at wish upon increasing the temperature,
always avoiding though to end in a melting regime for too large thermal fluctuations.

\section{Gaussian variational approach}\label{Sec_Variational}

In order to complement the FRG analysis we consider a Gaussian variational method (GVM) and compare the two methods.
Such a comparison in addition to providing some more transparent physical interpretation to the length scales is also of practical significance.
Although the FRG is essentially exact when $\epsilon \ll 1$ it become quantitatively unreliable in the interesting physical dimensions, and
is very difficult to extend to more complicated elastic terms. On the other hand the variational method has proven that it can also handle
such complications and thus can be used in more realistic situations to also compute the thermal crossover scales.

We follow here the methodology of Ref.~\cite{GLD95} and thus give only the main steps.

\subsection{Replicated hamiltonian}

The starting point of the Gaussian variational method is the following
replicated Hamiltonian~\cite{GLD95}:
\beq\label{Hreplica}
\begin{split}
H^n = & \frac{c}{2} \sum_a \int {\rm d}^d x (\nabla u^a(x))^2 \\
& - \frac{D}{2 T}\sum_{a,b} \int {\rm d}^d x \cos\left(K (u^a(x)-u^b(x)) \right) \ ,
\end{split}
\eeq
where $K$ is defined in Eq.~(\ref{Ru}).
We look for the best quadratic Hamiltonian, approximating~(\ref{Hreplica}):
\beq\label{H_GVM}
 H_0^n = \frac{1}{2} \sum_{a,b}  \int \frac{{\rm d}^d q}{(2 \pi)^d} G^{-1}_{ab}(q) u^a(q) u^b(-q) \ ,
\eeq
where $G_{ab}^{-1}$ is a $n\times n$ matrix of variational parameters.
We can choose $G_{ab}^{-1}$ of the form:
\beq
G_{ab}^{-1}(q) = c q^2 \delta_{a,b} - \sigma_{a b} \ ,
\eeq
where $\sigma_{ab}$ does not depend on $q$.
The matrix $G_{ab}^{-1}$ is found optimizing the variational free
energy $F_{var}^n = \langle H^n - H_0^n \rangle_0 + F_0^n$, where the average is
over $H_0^n$ and $F_0^n = - T \log H_0^n$.
We define the connected part as $G_c^{-1}(q) = \sum_b G^{-1}_{ab}(q)$.
From the minimization one obtains:
\beq
G^{-1}_{ab} = c q^2 \delta_{a,b} + \frac{1}{T} \frac{\partial}{\partial G_{a b}} \langle H_{dis}^n \rangle \ ,
\eeq
and the following saddle point equations follow:
\beq\label{Eqs_SP}
\begin{array}{cc}
\displaystyle \sigma_{a\neq b} =  \frac{D}{T} K^2 e^{- \frac{K^2}{2} B_{a b}(x=0)}
\\ \vspace{-0.2cm} \\
\displaystyle B_{ab}(x) = \langle [ u^a(x)-u^b(0) ]^2 \rangle
\\ \vspace{-0.2cm} \\
\displaystyle =
T \int \frac{{\rm d}^d q}{(2 \pi)^d} (G_{aa}(q) + G_{b b}(q) - 2 \cos (qx) G_{a b}(q))
\\ \vspace{-0.2cm} \\
\displaystyle\sigma_{a a} = - \sum_{b\neq a} \sigma_{a b} = \tilde{\sigma}
\\ \vspace{-0.2cm} \\
\displaystyle G_c(q) = \frac{1}{c q^2}
\ ,
 \end{array}
\eeq
which should be evaluated in the $n\to 0$ limit.
In the following we restrict  to the study of the full replica symmetry breaking (RSB)
ansatz of $G_{ab}$ as it is known to be the correct one~\cite{MP91,GLD95}.

\subsection{RSB ansatz}\label{GVM_RSB}

We parametrize the matrix $\widehat{G}$ with its diagonal terms $\tilde{G}$
and the off-diagonal terms by the function $G(u)$ with $u \in [0,1]$.
Similarly one has $\tilde{\sigma}$ and $\sigma(u)$. It is also convenient
to introduce the function:
\beq
[\sigma](u) = - \int_0^u {\rm d} v \, \sigma(v) + u \sigma(u) \ .
\eeq
We use the inversion formulas for hierarchical matrices defined in~\cite{MP91}.
The saddle point equations become:
\beq\label{Eqs_SP_RSB}
\begin{array}{cc}
\displaystyle \sigma(u) =  \frac{D}{T} K^2 e^{- \frac{K^2}{2} B(x=0,u)}
\\ \vspace{-0.2cm} \\
\displaystyle B(x=0,u) =
2 T \int \frac{{\rm d}^d q}{(2 \pi)^d} (\tilde{G}(q)  -   G(q,u)) \ .
 \end{array}
\eeq
We look for a solution of the form $\sigma(u)=const$ for $u>u_c$
and $\sigma(u)$ some function of $u$ for $u<u_c$, $u_c$ being itself
a variational parameter~\cite{GLD95}.
From the rules of inversion of algebraic matrices
we obtain:
\beq
\begin{array}{cc}
\displaystyle B(x=0,u) = & \displaystyle
2 T \int \frac{{\rm d}^d q}{(2 \pi)^d} \left[ \frac{1}{G_c^{-1}(q) + [\sigma](u_c)} \right.
\\ \vspace{-0.2cm} \\
& \displaystyle \qquad \left. + \int_u^{u_c} {\rm d} v
\frac{\sigma'(v)}{(G_c^{-1}+[\sigma](v))^2} \right] \ .
 \end{array}
\eeq
Taking derivative of (\ref{Eqs_SP_RSB}), beyond the solution $\sigma'(u)=0$,
in the limit $\sigma'(u)\neq 0$ one obtains:
\beq\label{Eq_SP_Der}
\begin{array}{cc}
\displaystyle 1 & \displaystyle = \sigma(u) \int \frac{{\rm d}^d q}{(2 \pi)^d} \frac{K^2 T}{(c q^2+[\sigma](u))^2}
\\ \vspace{-0.2cm} \\
& \displaystyle = \sigma(u)  \frac{T K^2 c_d}{c^{d/2}} ([\sigma](u))^{\frac{d-4}{2}}
 \end{array}
\eeq
with:
\beq
\begin{array}{cc}
\displaystyle c_d = \int \frac{{\rm d}^d q}{(2 \pi)^d} \frac{1}{(q^2+1)^2} & \displaystyle = \frac{(2-d)\pi^{1-d/2}}{2^{d+1}\sin(d\pi/2)\Gamma(d/2)}
\\ \vspace{-0.2cm} \\
& \displaystyle = \frac{(2-d)\pi S_d}{4\sin(d\pi/2)} \ .
 \end{array}
\eeq
Taking derivative of (\ref{Eq_SP_Der}) one gets:
\beq\label{Sol_S_u}
[\sigma](u) = \left(\frac{u}{u_0}\right)^{\frac{2}{d-2}} \ ,
\eeq
with $u_0 = 2 T K^2 c_d c^{-d/2}/(4-d) = T \tilde{u}_0$.
The solution (\ref{Sol_S_u}) is valid at small $u$ and we now determine
the breakpoint $u_c$ beyond which $[\sigma]=\Sigma$ is constant. We of course keep special
care in making the study for a finite $T$.
This can be done as follows. We write  (\ref{Sol_S_u}) as:
\beq
\displaystyle [\sigma](u) = \Sigma \left(\frac{u}{u_c}\right)^{\frac{2}{d-2}}
\eeq
with $u_c = u_0 \Sigma^{\frac{d-2}{2}}$.
From Eq.~(\ref{Eqs_SP_RSB}) and (\ref{Eq_SP_Der}) one finds:
\beq\label{Eq_Sigma}
\displaystyle \Sigma^{\frac{4-d}{2}} = \frac{D K^4 c_d}{c^{d/2}} e^{- \frac12 K^2 B(0,u_c)}
\eeq
with
\beq\label{Buc_RSB}
\displaystyle B(0,u_c) = 2 T  \int \frac{{\rm d}^d q}{(2 \pi)^d} \left[ \frac{1}{c q^2 +\Sigma} \right] \ .
\eeq

\subsection{Displacement correlation function with the GVM}\label{Sec_GVM_Sq}

One can now compute the roughness:
\beq
\begin{array}{ll}
\displaystyle B(x,0) & \displaystyle =  \overline{\langle (u_a(x)-u_a(0))^2 \rangle}
\\ \vspace{-0.2cm} \\
& \displaystyle =
 2 T  \int \frac{{\rm d}^d q}{(2 \pi)^d} (1 - \cos(q x)) \tilde{G}(q)
  \end{array}
\eeq
with:
\beq\label{Eq_Gamma_GVM}
\begin{array}{ll}
\displaystyle \tilde{G}(q) & \displaystyle =
\tilde{G}_{th}(q) + \tilde{G}_{dis}(q)
\\ \vspace{-0.2cm} \\
& \displaystyle = \frac{1}{c q^2} + \frac{1}{c q^2} \int_0^1 \frac{{\rm d} v }{v^2} \frac{[\sigma](v)}{c q^2+[\sigma](v)} \ ,
   \end{array}
\eeq
where we have used the rules of inversion of hierarchical matrices~\cite{MP91} and that
$\sigma(0)=0$.
Therefore:
\beq
 \displaystyle B(x) = B_{th}(x) + B_{dis}(x) \ ,
\eeq
where:
\beq
\displaystyle B_{th}(x) = 2 T  \int \frac{{\rm d}^d q}{(2 \pi)^d} (1 - \cos(q x)) \frac{1}{c q^2}
\eeq
is the thermal part of a non disordered system and:
\beq
\begin{array}{ll}
\displaystyle B_{dis}(x) &
 \displaystyle  =  2 T  \int \frac{{\rm d}^d q}{(2 \pi)^d} (1 - \cos(q x)) \times
\\ \vspace{-0.2cm} \\
& \displaystyle \qquad\qquad\qquad \times\frac{1}{c q^2}\int_0^1 \frac{{\rm d} v }{v^2}  \frac{[\sigma](v)}{c q^2+[\sigma](v)}
\\ \vspace{-0.2cm} \\
& \displaystyle = 2 T  \int \frac{{\rm d}^d q}{(2 \pi)^d} (1 - \cos(q x)) \times
\\ \vspace{-0.2cm} \\
&\displaystyle \qquad \times
\left[ I(q)
+\frac{1}{c q^2} (u_c^{-1}-1) \frac{\Sigma}{c  q^2+\Sigma}
\right] \ .
 \end{array}
\eeq
We want to deal with the integral:
\beq
\begin{array}{ll}
 \displaystyle I(q) &  \displaystyle =  \frac{1}{c q^2}  \int_0^{u_c} \frac{{\rm d} v }{v^2}  \frac{v^{\mu}}{u_0^{\mu} c q^2+v^{\mu}}
 \end{array}
\eeq
where we define $\mu= \frac{2}{d-2}$. With $q_0=\sqrt{\frac{1}{c} (\frac{u_c}{u_0})^{\mu}} = \sqrt{\Sigma/c}$
 the limit $q/q_0 \ll 1$ reads:
\beq
\begin{array}{ll}
 \displaystyle I(q)  \displaystyle \stackrel{q/q_0\ll1}{\simeq}  \frac{1}{q^{d}} Y - \frac{1}{u_c c q^2} + {\mathcal O}(1)
\ ,
 \end{array}
\eeq
with $Y=\frac{(d-2) \pi}{2 u_0 \sin(\frac{\pi (d-2)}{2})}c^{-\frac{d}{2}}$,
which gives the desired power law behavior $q^{-d}$ at large distances.
While in the limit $q/q_0\gg 1$ one gets:
\beq
\begin{array}{ll}
 \displaystyle I(q)  \displaystyle \stackrel{q/q_0\gg1}{\simeq}  \displaystyle
\frac{1}{c q^2 u_c}  \left[ \frac{1}{\mu-1} \frac{q_0^2}{q^2} + {\mathcal O}\left(\frac{q_0^4}{q^4}\right) \right] \ .
 \end{array}
\eeq

We see that the correlation function is made by a thermal part:
\beq
\tilde{G}_{th}(q) = 1/(c q^2) \ ,
\eeq
a term corresponding to a modified Larkin regime:
\beq
\begin{array}{ll}
\displaystyle  \tilde{G}_{L}(q)
& \displaystyle   \stackrel{q/q_0\gg1}{\simeq} \frac{1}{c q^2} \frac{1}{u_c}\frac{\Sigma}{c q^2} [  \frac{\mu}{\mu-1} - u_c ]
\\ \vspace{-0.2cm} \\
& \displaystyle =  \frac{1}{c q^2} \frac{1}{u_c}\frac{\Sigma}{c q^2} [  \frac{2}{4-d} - u_c ]
\ ,
\end{array}
\eeq
and for large distances
the term which gives logarithmic growth:
\beq
\tilde{G}_{RP}(q) \stackrel{q/q_0\ll1}{=} \frac{1}{q^d} Y  \ .
\eeq

\subsection{Crossover length scales within the GVM}\label{Sec_GVM_lengths}

The previous expressions allow us to extract the crossover scales within the GVM.
We define $l_{L}$ and correspondingly $q_{L}=l^{-1}_{L}$
the length such that $q_{L}=q_0=\sqrt{\Sigma/c}$ which corresponds to the
region of validity of the Larkin regime.
Assuming $l_{L}\gg a$ one has $B(0,u_c) \simeq L_{l}^2$ where $L_{l}$ is the Lindemann length defined
in (\ref{lT}) and  one has from (\ref{Eq_Sigma}):
\beq\label{qL_GVM}
\displaystyle q_{L} = \sqrt{\frac{\Sigma}{c}} = \frac1a \Big(\frac{\tilde{D} (Ka)^4 c_d}{\tilde{c}^2} \Big)^{1/\epsilon} e^{- L_{l}^2 K^2/2\epsilon}
\eeq
and:
\beq
\displaystyle u_c =  \frac{2 T K^2 c_d }{(4-d) c } q_{L}^{d-2} \ .
\eeq

In the limit $T\to 0$ the breakpoint $u_c$ goes to $0$ but $q_{L}$ remains finite.
In the limit of high temperature instead $q_{L}\to 0$ and also $u_c$.
Similarly to what has been done with the results obtained by FRG,
the crossover between the thermal and the Larkin regime can be determined by the condition:
\beq
\displaystyle \tilde{G}_{th}(q_{th}) =  \tilde{G}_L(q_{th}) \ .
\eeq
This gives:
\beq\label{qth_def3}
\begin{array}{ll}
\displaystyle q_{th} & = \displaystyle \sqrt{\frac{\Sigma}{c}  \frac{1}{u_c} } \sqrt{ \frac{2}{4-d} - u_c}
\\ \vspace{-0.2cm} \\
& \displaystyle = \sqrt{ \frac{ (4-d) K^2 D}{2 c T}} e^{-L_{l}^2 K^2/4}  \sqrt{ \frac{2}{4-d} - u_c}
\\ \vspace{-0.2cm} \\
& \displaystyle
\simeq K \sqrt{ \frac{  \tilde{D}}{\tilde{c} T}} e^{-l_T^2 K^2/4}
\ .
\end{array}
\eeq
Roughly with this definition one has $q_{th}> q_{L}$ as far as $u_c \epsilon/2<1$ and $q_{th}= q_{L}$ for $u_c \epsilon/2 \simeq 1$.
This condition might be violated at intermediate temperatures if disorder
is sufficiently high leading to the disappearance of the intermediate
Larkin regime.

\section{Comparison between FRG and GVM and discussion}\label{Sec_Sq_Comparison}

\begin{figure}
  \begin{center}
    \includegraphics[height=6cm]{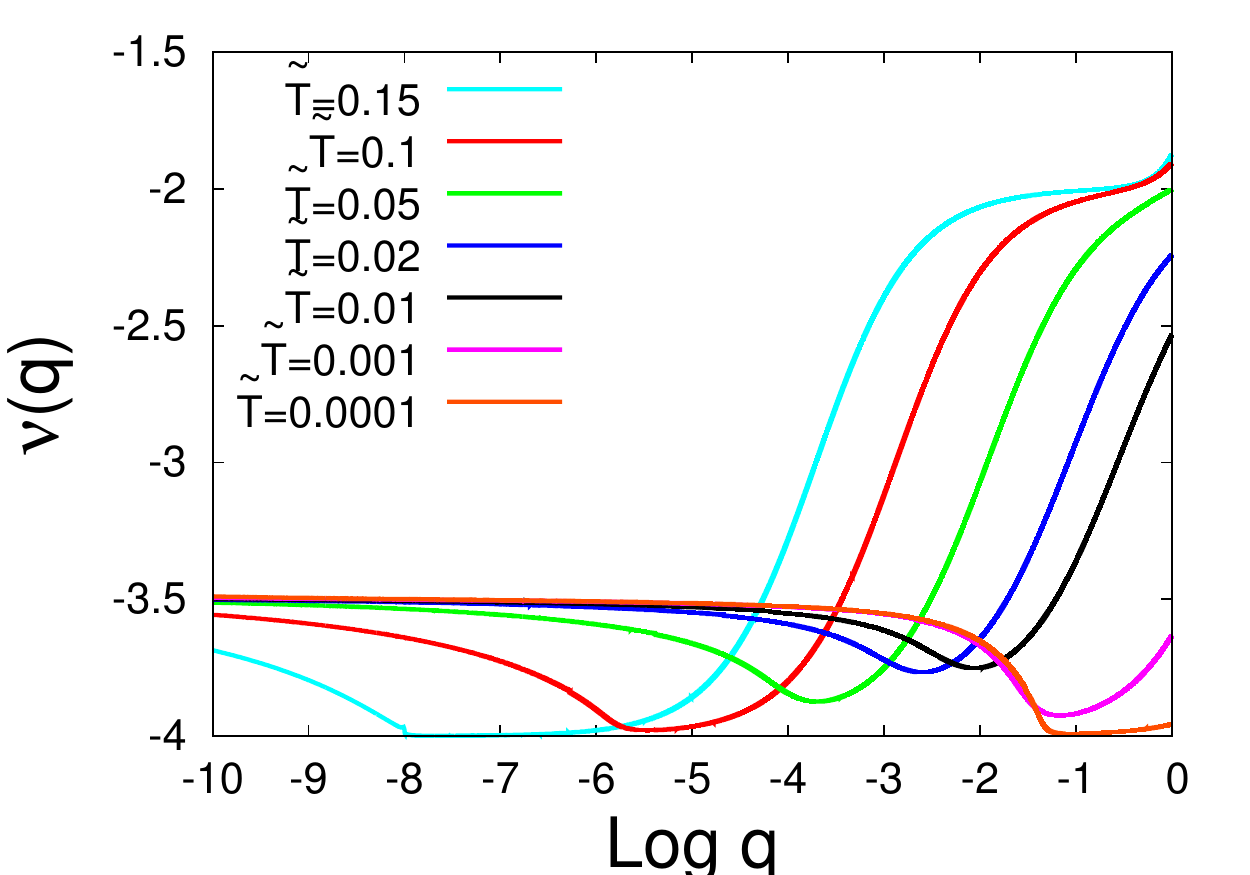}
        \caption{Logarithmic derivative of $\Gamma(q)$ obtained by FRG for different temperatures.
	        In particular we show $\tilde{T}=0.15$, $\tilde{T}=0.1$, $\tilde{T}=0.05$, $\tilde{T}=0.02$, $\tilde{T}=0.01$, $\tilde{T}=0.001$ and $\tilde{T}=0.0001$
        respectively with light blue, red, green, blue, black, pink and orange solid lines.
        The other parameters are $\epsilon=0.5$ and $c=1$. $\tilde{\Delta}_{l=0}(0)=0.005$.
}
    \label{Fig:LogDer_FRG}
  \end{center}
\end{figure}

\begin{figure}
  \begin{center}
    \includegraphics[height=6cm]{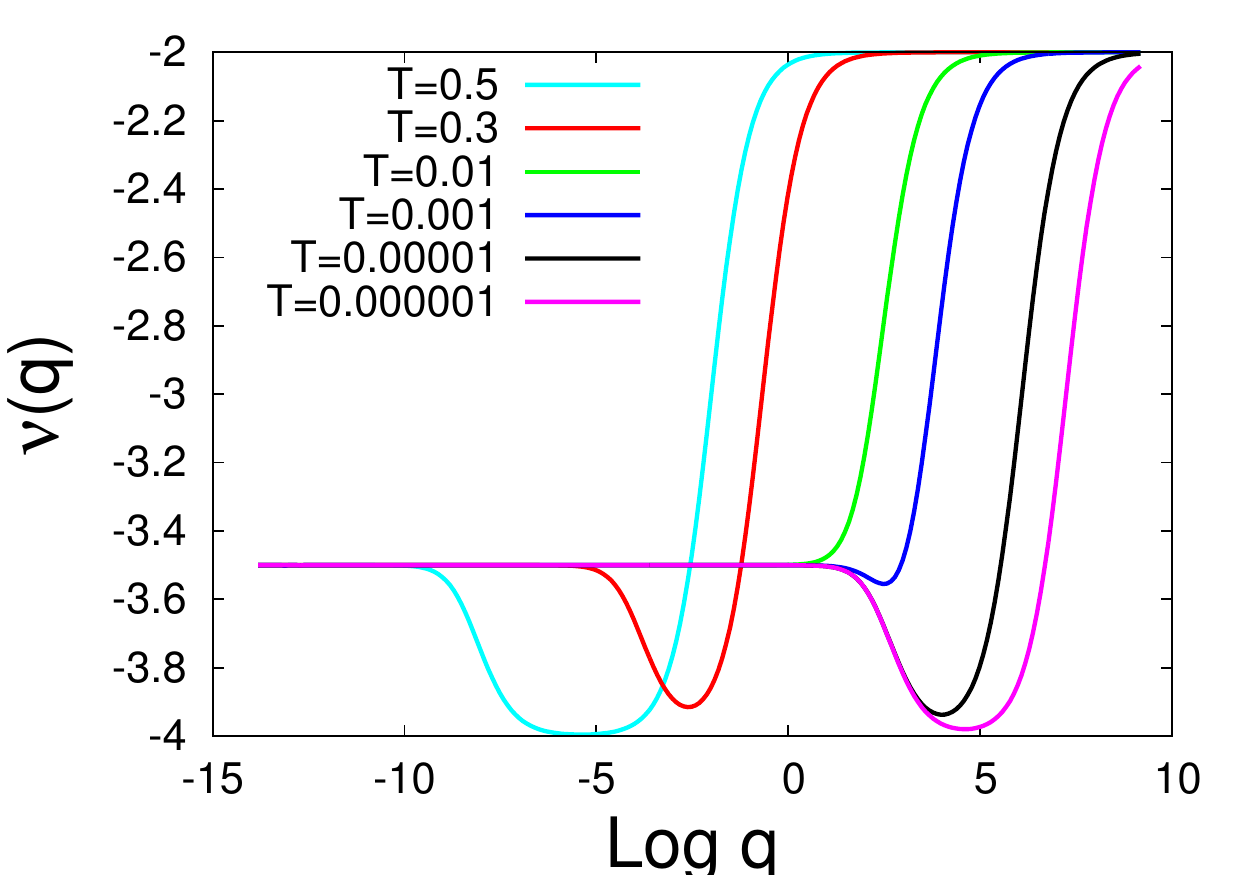}
        \caption{
        Logarithmic derivative of $\Gamma(q)$ obtained by GVM.
        The temperatures shown here are $T=0.5$, $T=0.3$, $T=0.01$, $T=0.001$, $T=0.00001$ and $T=0.000001$,
        respectively with light blue, red, green, blue, black and pink solid lines. The other parameters have been fixed
       to
        $c=1$, $a=1$, $\epsilon=0.5$ and $D=0.05$.
        With this choice of parameters one sees that the intermediate Larkin regime,
        at intermediate temperatures, tends to disappear.
Moreover one also sees that at low temperatures the inverse length scale $q_L$, namely the inverse length scale associated to the passage from
the Larkin to the random periodic regime, saturates to its $T=0$ value, while $q_{th}$ is
pushed towards larger and larger values as $T$ goes to zero. At high temperatures
$q_L$ and $q_{th}$ are sent to smaller and smaller values with $q_L < q_{th}$.
}
    \label{Fig:LogDer_GVM_D0_0.05}
  \end{center}
\end{figure}

In this section we discute and compare the results obtained by FRG and by GVM.

All the results are valid in the elastic limit $u_{i+1}-u_i \ll a$ and concerning the FRG they are expected to
be accurate at small $\epsilon$ where $\epsilon=4-d$.
In Figs.~\ref{Fig:LogDer_FRG} and \ref{Fig:LogDer_GVM_D0_0.05}
we show the logarithmic derivative
of the full solution of the displacement correlation function $\nu(q) = \frac{{\rm d} \log \Gamma_q}{{\rm d} \log q}$
for different temperatures that highlights the different regimes and
the associated exponents.
Fig.~\ref{Fig:LogDer_FRG} is obtained by FRG, according to Eq.~(\ref{Eq_Gamma}),
while Fig.~\ref{Fig:LogDer_GVM_D0_0.05} is the result of the GVM, i.e. Eq.~(\ref{Eq_Gamma_GVM}).
In both cases we have considered $\epsilon=0.5$.

The two figures clearly show that at high and low temperatures three regimes, characterized by three different exponents, are present: the thermal regime with $\nu = 2$, the Larkin regime with $\nu=4$
and the asymptotic random periodic with $\nu=d$, which for the parameters used here is $\nu=3.5$.
Correspondingly the roughness exponent is $\zeta=(\nu-d)/2$, and it is associated to logarithmic grow of the displacements when $\nu=d$.

As can be seen both from the figures and also from the analytical estimates both methods are in remarkable agreement for the crossover scales. In particular
the crossover inverse length scale between the thermal and the Larkin regime is given in Eq.~(\ref{qth_FRG})
and (\ref{qth_def3}) while the one between the Larkin and the random periodic is~(\ref{qL_FRG})
and (\ref{qL_GVM})
(in the limit of small $\epsilon$ the
quantity $c_d$ appearing in (\ref{qL_GVM}) goes as $c_d= S_d/\epsilon$).
Note that, apart via the Lindemann length, the quantity $q_{th}$ does not depend on
the dimension of the system, contrarily to $q_L$ which does directly depend on $\epsilon$.
The disorder strength instead appears explicitly in both expressions.
As clear from both the FRG and GVM study below the scale $l_{th}$ the disorder is essentially absent and the system behaves like a pure thermal system.

These two length scales have very different behavior at low and high temperature.
In both cases at high enough temperatures the Lindemann length intervenes
in an exponential way in the corresponding length scale traducing the exponential screening of the disorder by the thermal fluctuations. This is visible on Figs.~\ref{Fig:LogDer_FRG} and \ref{Fig:LogDer_GVM_D0_0.05} which confirms
that the two inverse length scales are sent towards smaller and smaller values with $q_{th}>q_L$.
Note that this high temperature limit is only valid with systems for which the elastic limit can be enforced even if the temperature is high, such as e.g. the system of lines of Fig.~\ref{Fig:lines}. In point like solids topological defect will be induced by the temperature and the solid will melt when the Lindemann length equals $L_l \sim C_l a$ where $C_l \sim 0.1$ (Lindemann criterion of melting).

At low temperature $L_l \ll a$ the exponential factor plays little role.
This implies that $q_L$ becomes essentially temperature independent at low temperature in agreement with the fact that the problem is asymptotically determined by the zero temperature fixed point, when the disorder is small. Of course for finite disorder the full solution of the flow is needed and some residual if weak temperature dependence will be present in the scale $q_L$.
This qualitative behavior is clearly visible in the full solution in
Figs.~\ref{Fig:LogDer_FRG} and \ref{Fig:LogDer_GVM_D0_0.05} where
one sees that all the curves at low enough temperature
overlap in the crossover region around $q_L$.
On the contrary the crossover inverse length $q_{th}$ between the
thermal and the Larkin regime is pushed towards larger and larger values as $T$ goes to zero.

For intermediate temperature the figures~\ref{Fig:LogDer_FRG}  and~\ref{Fig:LogDer_GVM_D0_0.05} show that the Larkin regime tends to disappear
in agreement (for high enough disorder strength)
with the analysis carried on within the FRG and the GVM.
This regime of temperatures is shown with a blue and a black line in Fig.~\ref{Fig:LogDer_FRG} and with a bue and a green line in Fig.~\ref{Fig:LogDer_GVM_D0_0.05}.

\begin{figure}
  \begin{center}
    \includegraphics[height=6cm]{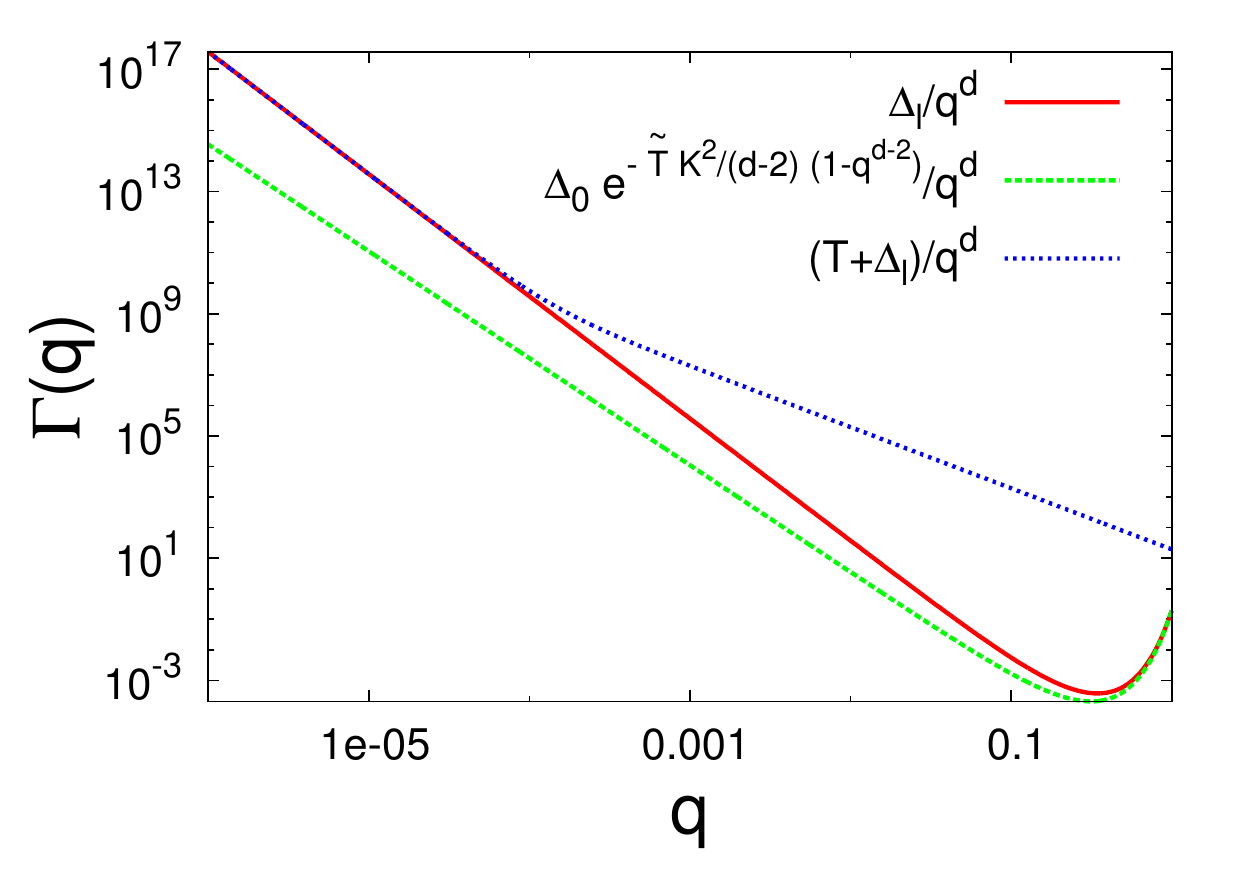}
        \caption{
       The solid red line indicates the disordered part of the FTD which shows
       at large momenta an unexpected increasing behavior upon increasing $q$. The dashed
       blue line is the full FTD as in Eq.~(\ref{Eq_Gamma}) where the thermal
       part washes out the non-monotonicity. The dashed green line is the result as obtained from the flow
       when only the linear term proportional to temperature is kept. In this plot $\epsilon=0.5$ and
       $\tilde{T}=0.5$. From these value one obtains $q_T\simeq 0.16$.
       }
    \label{Fig:Correction_Sq}
  \end{center}
\end{figure}

We finally mention that within the FRG we find an additional length scale that is not present within the GVM.
Such scale is given in Eq.~(\ref{qT_FRG})
and does depend on temperature and on the dimension of the system but it is independent of the disorder strength. It corresponds to a flow of the correlator towards a vanishing
amplitude function (see the lower panel of Fig.~\ref{Fig:FRG_T_v0}).
It would be interesting to test experimentally if such length scale can be observed.
However this length scale does not show in an obvious way in the displacement correlation function since it is not accompanied by a change of exponent.  This is due to the fact that at that corresponding length scale the system is dominated by the thermal part of the displacement and that such a length scale only affects the ``disorder'' part of the correlation.
In order to observe it it is necessary as can be seen from Fig.~\ref{Fig:Correction_Sq} to subtract
the thermal part.
In particular, if one keeps only the term proportional to temperature in the flow,
the disordered part of the FTD for $ q = e^{-l}$ reads
$\Gamma(q,T,\Delta) = \Big(\frac{1}{q}\Big)^d  \Delta_0 e^{- \frac{\tilde{T} K^2}{d-2} (1-q^{d-2})} $.
At high temperature, in the regime $q>q_T$, this correction of the FTD to the
thermal part results in an unexpected behavior that \emph{decreases} upon \emph{decreasing} of $q$ (see Fig.~\ref{Fig:Correction_Sq}). This corresponds to a screening of the disorder by the thermal fluctuations leading to a reduced disorder as one looks at larger and larger length scales.
Note that although this length scale is always present in our purely elastic model, it is even more subject to the constraints on the high temperature limit in a model where melting (i.e. the presence of topological defects can occur) than the two other length scales. 
Indeed (\ref{qT_FRG}) can be written for small $\epsilon$ as 
\begin{equation} 
 q_T / \Lambda \simeq (\Lambda a)^{\frac{d}2-3} \frac{\epsilon^{1/2}}{L_{l}/a}
\end{equation}
where we have assumed $\Lambda \sim K$. One can estimate $\Lambda a \sim \pi$ and 
$L_l / a \sim C \sim 0.1$ if the melting can occur. In that case one would need a system which is effectively close to four dimensions (with e.g. long range elastic couplings 
such as in ferroelectrics~\cite{LK69}). On the contrary in the model of lines of Fig.~\ref{Fig:lines} the inverse length scale $q_T$ should be visible in particular if the temperature is high enough.

\section{Conclusions}\label{Sec_concl}

We have considered a system described by an elastic Hamiltonian
and subject to a disordered environment with periodic correlation functions,
as it could be for charge density waves.
We have analyzed the system by functional renormalization group techniques
and a Gaussian variational approach.
Both approaches can be applied to arbitrary dimensions even if
the FRG is believed to be accurate around $4 - \epsilon$ dimensions.
Within these two methods we have computed the
relative displacement correlation and its logarithmic derivative taking into account the effects of a finite temperature.
To do so we have introduced an approximation scheme for the FRG equations which is quite generic and can be used in more complex situations such as the dynamics.

We find three regimes as a function of the wavevector (or in real space
the distance) for which the Fourier transform of the displacement correlation function (FTD)
behaves essentially with a power law of the wavevector characterized by different
exponents in each regime: the thermal, Larkin and random periodic regimes. In the first regime the system behaves as a pure
elastic system at finite temperature. In the second (Larkin) the system sees a disorder which is essentially like a random force,
while in the asymptotic and last regime the periodicity plays a full role and  leads to a logarithmic growth of the correlations in real space.
For each transition from one regime to the other we have
determined the crossover length scale as a function of the parameters
defining the model, and in particular the temperature. Both the FRG and GVM give consistent results on these two length scales.

At large temperatures, in an ideal elastic systems these two scales would grow exponentially with the Lindemann length of the systems.
In practice one should of course worry about the melting of the corresponding periodic system.

At low temperatures the thermal regime
length scale goes to zero while the length scale separating the Larkin and asymptotic regimes stays finite, consistently with previous results.
At intermediate temperatures depending on the parameters it is possible to remove the Larkin regime and to have
a direct transition between the thermal and random periodic (Bragg glass) regime.
Besides these three regimes  we find by FRG an
additional length scale which characterizes the FTD once the thermal part
is subtracted. Within such length scale and at high enough temperature one finds that the disordered part of the FTD
has a non monotonic behavior with $q$, as shown in Fig.~\ref{Fig:Correction_Sq}.

It would of course be interesting to check if the predicted temperature of the length scales computed here can be observed in experiments or simulations.
In particular finding evidence of the scale $q_T$ of Eq.~(\ref{qT_FRG})
by measuring the relative displacement correlation and subtracting the thermal part should prove interesting.

As a future perspective it is of interest to see how these
crossover length scales and the FTD are modified
by the influence of a finite velocity which is present in the driven system at finite temperature. In particular for such a study it would be interesting to see whether the fitting form that is found here can be used to replace the
functional RG equation with standard RG equations for the parameters involved in the fitting form.

\acknowledgments
We thank Elisabeth Agoritsas and Vivien Lecomte for valuable discussions.

\bibliographystyle{mioaps}
\bibliography{DES}

\end{document}